\documentclass[conference]{IEEEtran}
\IEEEoverridecommandlockouts

\usepackage{threeparttable}
\usepackage{booktabs}
\usepackage{cite}
\usepackage{amsmath,amssymb,amsfonts}
\usepackage{algorithmic}
\usepackage{graphicx}
\usepackage{textcomp}
\usepackage{multirow}
\usepackage{float}

\usepackage[numbers]{natbib}
\usepackage{hyperref}

\def\BibTeX{{\rm B\kern-.05em{\sc i\kern-.025em b}\kern-.08em
    T\kern-.1667em\lower.7ex\hbox{E}\kern-.125emX}}

\usepackage[table]{xcolor}

\definecolor{custom-gray}{cmyk}{0, 0, 0, 0.7, 1.00}
\usepackage[most]{tcolorbox}
\usepackage{cleveref}
\newtcbtheorem[no counter]{Summary}{\hskip-0.97em}{enhanced,drop shadow={black!50!white},
 coltitle=white,
 top=0.15in,
 separator sign none,
 attach boxed title to top left=
 {xshift=0em,yshift=-\tcboxedtitleheight/2},
 boxed title style={size=small,colback=custom-gray}
}{summary}

\begin{document}

\title{An Empirical Study on the Capability of LLMs in Decomposing Bug Reports
}


\author{Zhiyuan Chen$^{1}$, Vanessa Nava-Camal$^{1}$, Ahmad Suleiman$^{1}$, Yiming Tang$^{1,*}$, Daqing Hou$^{1}$ and Weiyi Shang$^{2}$\\
	\normalsize $^{1}$Rochester Institute of Technology, Rochester, New York, United States\\
	\normalsize $^{2}$University of Waterloo, Waterloo, Ontraio, Canada\\
	\normalsize \{zc9482, vn4261, as4300, yxtvse, dqvse\}@rit.edu$^{1}$,  {wshang}@uwaterloo.ca$^{2}$\\
	\normalsize *corresponding author
}

\maketitle

\begin{abstract}

\noindent \textit{\underline{Background:}} Bug reports are essential to the software development life cycle. They help developers track and resolve issues, but are often difficult to process due to their complexity, which can delay resolution and affect software quality. 

\noindent \textit{\underline{Aims:}} This study investigates whether large language models (LLMs) can assist developers in automatically decomposing complex bug reports into smaller, self-contained units, making them easier to understand and address.

\noindent \textit{\underline{Method:}} We conducted an empirical study on 127 resolved privacy-related bug reports collected from Apache Jira. We evaluated ChatGPT and DeepSeek using different prompting strategies.
We first tested both LLMs with zero-shot prompts, then applied improved prompts with demonstrations (using few-shot prompting) to measure their abilities in bug decomposition.

\noindent \textit{\underline{Results:}} Our findings show that LLMs are capable of decomposing bug reports, but their overall performance still requires further improvement and strongly depends on the quality of the prompts. With zero-shot prompts, both studied LLMs (ChatGPT and DeepSeek) performed poorly. After prompt tuning, ChatGPT's true decomposition rate increased by 140\% and DeepSeek's by 163.64\%.

\noindent \textit{\underline{Conclusions:}} LLMs show potential in helping developers analyze and decompose complex bug reports, but they still need improvement in terms of accuracy and bug understanding.



\end{abstract}


\section{Introduction}


Bug reports are a critical resource in the software development lifecycle (SDLC), providing essential information about issues for debugging, testing, and maintenance \cite{PhilipJ2011}. 
Bug reports typically contain descriptions of software bugs, such as the location where the fault occurred, which helps developers quickly locate and resolve issues.
By offering timely insights into software issues, bug reports facilitate reducing development costs and improving software quality efficiently.

Despite their importance in software development, 
bug reports are often too complex to support timely resolution. As modern software systems grow in scale and sophistication, the bugs they generate also become more complex~\cite{Catolino2019}, resulting in increasingly detailed and multifaceted bug reports. These reports may include diverse types of information such as crash descriptions, reproduction steps, and test cases~\cite{Soltani2020}, which requires manual effort from developers to analyze. Additionally, the length of bug reports can vary greatly, sometimes reaching several thousand lines~\cite{Mani2012}, which further increases the difficulty of manual review and understanding.





To help developers better understand bug reports, researchers have proposed various automated techniques for bug report analysis. For example, \citet{Steven2014} developed methods to extract key information from bug reports, such as error types and affected components, using natural language processing techniques. 
However, these methods often struggle due to the diversity of programming languages used in bug reports, as well as the lack of standardized formats for consistent and reliable analysis.
\citet{John2006} proposed a semi-automated approach to assign bug reports to the most appropriate developers based on their expertise and historical data. While effective, this approach relies heavily on the quality of the initial bug report.
\citet{Nicolas2008} identified duplicate reports and categorized them as bug reports, and \citet{Yuan2012} built on this by improving the accuracy of automated identification of duplicate bug reports.
Additionally, some studies have explored text summarization \cite{Sarah2014} or classification \cite{zhou2016combining} methods to extract useful information from bug reports. 
While these studies offer valuable insights to facilitate bug analysis, they always overlook the bug complexity in bug reports. 
Rather than being atomic, bug reports often consist of multiple units, each of which could be treated as a separate bug assigned to developers. Such complexity could reduce developers' motivation to resolve bugs.  
Exploring approaches to decomposing bug reports into smaller, self-contained units may significantly ease the resolution process and reduce developer effort.




In this paper, we propose a novel approach to bug report decomposition using LLMs to simplify bug reports and facilitate bug resolution. Recent advancements in LLMs such as ChatGPT have demonstrated impressive capabilities in understanding and generating human-like text across a wide range of domains \cite{antoun2024textsourceresultsdetecting, kalyan2023surveygpt3familylarge, brown2020languagemodelsfewshotlearners}. Their ability to reason over complex natural language input makes them well suited for analyzing bug reports, which often lack formal structure and include various technical expressions, requiring much stronger processing capability to properly handle. 


Our empirical results show that LLMs' performance in bug report decomposition depends heavily on the quality of the prompt design. Both ChatGPT and DeepSeek struggled when given a plain prompt, correctly decomposing only 10 and 11 out of 127 bug reports, respectively. After applying few-shot prompting with decomposition-specific examples, ChatGPT and DeepSeek reached 240\% and 263.64\%  of their original performance, respectively, but their overall performance still has room for improvement. We identified four major causes of incorrect bug decompositions: \textit{over-decomposition}, \textit{over-analysis}, \textit{incorrect interpretation of solutions}, and \textit{lacking key information}. 
Our results may serve as inspiration for future research in automated bug analysis.


The contributions of this study can be summarized as follows:

\begin{itemize}
    \item  We propose an LLM-based approach to automatically decomposing complex bug reports into smaller, self-contained, well-structured units, each of which can be independently assigned to developers. Our replication package can be found at this link\footnote{\url{https://github.com/LLMPrivacyResearch/LLMBugReport}}.
    \item This study identifies four major causes of incorrect bug decomposition by LLMs.
    \item This study offers useful suggestions to bug reporters to facilitate automated bug decomposition and bug analysis.
\end{itemize}

\noindent \textbf{Paper Organization.} 
The remainder of this paper is organized as follows. Section~\ref{sec:background} discusses background and related work on bug report analysis. Section~\ref{sec:overview} introduces the study overview and 
Section~\ref{sec:results} describes the study results. Section~\ref{sec:threat} presents threats to validity and discusses the limitations of our study. Finally, Section~\ref{sec:conclusion} concludes the paper and outlines future directions.

\section{Background and related work} \label{sec:background}

\textbf{Bug reports play a significant role in SDLC.}
A bug report typically includes a documented description of the bug, steps to reproduce the issue, details about the execution environment, and, in some cases, code snippets or logs that support the bug diagnosis and resolution process~\cite{Steven2014}.
As the primary means by which software users communicate problems to developers~\cite{Davies2014}, bug reports are important for key phases of SDLC, such as testing and maintenance.
With the emergence of large-scale software systems, maintenance activities have increasingly relied on bug reports~\cite{Zhang2016}.
An existing study~\cite{Bob2008} reveals that software maintenance accounts for 60\% of the total cost of SDLC, which indicates that effective bug report analysis and resolution can significantly reduce development costs.
A survey~\cite{Soltani2020} conducted on 305 developers demonstrates that the information contained in bug reports is crucial for software debugging, particularly crash descriptions, reproduction steps or test cases, and stack traces.

\textbf{The complexity of bug reports prevents timely bug resolution.}
Bug report analysis usually requires significant manual effort, as developers have to manually analyze suggested bug content, which may vary in length from a few lines to several thousand lines~\cite{Mani2012}. 
Bugs evolve over time, becoming increasingly complex~\cite{Catolino2019}, which in turn contributes to the increasing complexity of bug reports and makes them more difficult for developers to analyze.
Bug reports exhibit diverse levels of complexity. A survey~\cite{Bettenburg2007} that collected 14 comments from various developers revealed that bug reports have different knowledge levels and involve complicated steps to reproduce, making it difficult for developers to analyze them.
\citet{Herraiz2008} point out that Eclipse's bug reports are too complex, which eventually affects the accuracy of developers' prioritization of bug reports.
Despite existing work uncovering the complexity of bug reports from multiple perspectives, there is currently no study that reveals the status of multiple bugs within a bug report or explores methods for simplifying them by decomposing them into smaller, self-contained units to facilitate bug resolution.




\textbf{With the emergence of LLMs assisting in software engineering tasks, there is still a lack of understanding of how LLMs decompose bug reports to facilitate automated bug report analysis.}
In recent years, LLMs, represented by ChatGPT~\cite{Liu2023} and DeepSeek \cite{xin2024deepseekproveradvancingtheoremproving}, are profoundly changing software engineering practices. Studies \cite{Xinyi2024} have shown that these models have reached human levels in tasks such as code generation like GitHub Copilot and automated bug fixing like DeepDebug \cite{drain2021deepdebugfixingpythonbugs}. Unlike earlier rule-based systems \cite{Vertsel2024}, LLMs are able to understand the informal expressions of developers by analyzing massive amounts of data \cite{MiaMohammad2024} from platforms such as Stack Overflow and GitHub. This capability is particularly useful for unstructured text such as bug reports, where reporters often use colloquial descriptions rather than standard technical language \cite{NEURIPS2024_71c3451f}.
LLMs can handle complex, multi-step instructions by reasoning through them step-by-step when appropriately prompted \cite{wei2023chainofthoughtpromptingelicitsreasoning}.
LLM-based systems can even decompose and orchestrate tasks: for instance, an agent powered by ChatGPT can take a single complex request and break it into sub-tasks \cite{shen2023hugginggptsolvingaitasks}.
While the application of LLMs in software engineering is gaining increasing attention, to the best of our knowledge, there is still a lack of studies that systematically examine their capabilities in bug decomposition.

\section{Case Study Overview} \label{sec:overview}

\subsection{Methodology Overview} \label{sec:rq}

\begin{figure*}[h]
    \centering
    \includegraphics[width=0.8\textwidth]{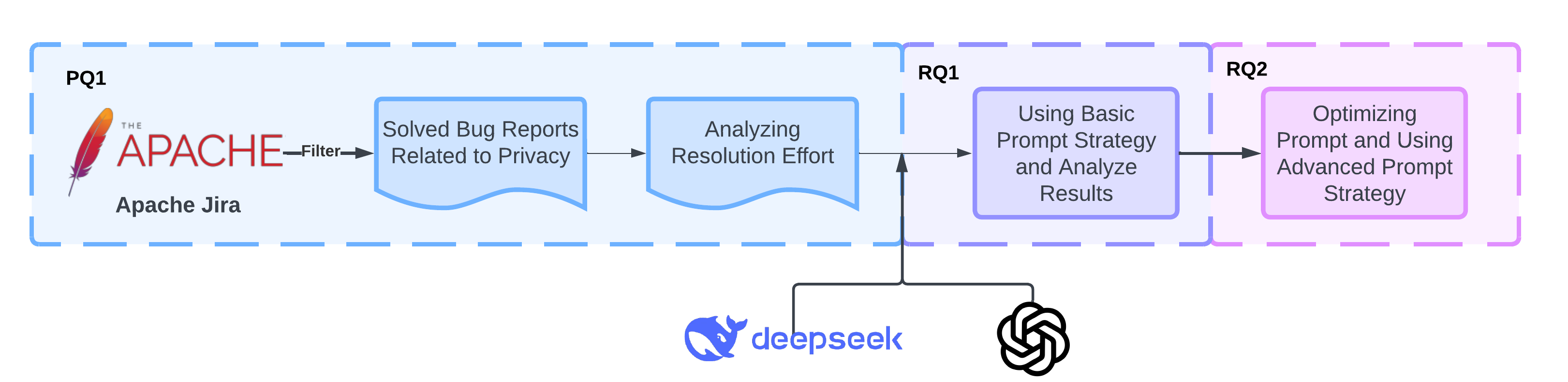}
	\caption{Workflow of studying decomposition capability of LLMs.} 
	\label{fig:workflow_overall} 
 \end{figure*}

To systematically evaluate the effectiveness of the LLMs to decomposing bug reports, we divided our study into three parts: one preliminary question and two research questions, as illustrated in Figure~\ref{fig:workflow_overall}. 
In PQ1, we explore the developers' efforts involved in bug report resolution, measured by resolution time, to reveal the current challenges in bug report resolution. In RQ1 and RQ2, we use two different prompting strategies, zero-shot and few-shot prompting, to explore the abilities of LLMs in bug decomposition. 

\subsection{Case Study Setup} \label{sec:setup}

In this section, we present how we collect bug reports for analysis and how we select LLMs.

\paragraph{Bug report collection} 

We use Apache Jira\footnote{\url{https://issues.apache.org/jira/secure/Dashboard.jspa}} to collect bug reports. As one of the largest bug report repositories, Apache Jira covers over 300 open-source projects from the Apache Software Foundation \cite{Apache_Jira_Software_2024}. It has been widely used in many studies on bug analysis \cite{Almeida2022,Veena2020}.

Since there are hundreds of bug reports on Apache Jira, it is not feasible to study all of them. Therefore, we decided to focus on a specific domain—one that has a significant impact on software quality, provides a sufficient number of bug reports without overwhelming developers, and is not tied to particular programming languages or techniques that might introduce bias into our analysis. Additionally, we sought a domain where the bugs are not easy to fix. After careful consideration, we chose to focus on privacy-related bug reports to analyze their complexity and explore the possibility of using LLMs to simplify their resolutions.

We used ``privacy'' as the keyword for searching. Since our study also involves measuring developers' efforts in bug resolution, we focused only on resolved bug reports. However, limiting the search to resolved bug reports initially yielded only 77 reports. To expand our dataset, we applied an additional search criterion that included bug reports not explicitly labeled as ``resolved'' but whose resolution status was marked as ``fixed.''
As a result, we collected 128 privacy-related bug reports. However, since one of them lacked a bug description, we excluded it, resulting in a final dataset of 127 valid bug reports.


\input{results.texi}

\subsection*{\textbf{PQ: How much time and effort do developers take to resolve bug reports?}}

\subsubsection*{\textbf{Motivation}}


Bug reports are crucial to software development and maintenance, as they document issues that need to be addressed to ensure software quality and reliability. However, resolving these reports often can be time-consuming and labor-intensive, which typically involves several steps, including understanding the problem, reproducing the problem, determining the root cause, implementing the fix, and validating the solution~\cite{John2006}. Given this complexity, manual effort plays a significant role in the bug resolution process.



Since this study focuses on the potential for automating the simplification of bug reports, it is essential to first understand whether, in practice, resolving bug reports actually requires significant developer effort. Therefore, this PQ aims to analyze how much time and effort developers typically spend on resolving bug reports.


\subsubsection*{\textbf{Approach}}

We use a straightforward metric—the time required to resolve bug reports—to measure the manual effort developers invest in bug resolution.
Specifically, we examined the time elapsed from the creation time of the bug report to its resolution time. However, this metric alone may not fully capture the actual time developers spent resolving the bug. This is because the interval between creation time and resolution time does not necessarily reflect when the developer actively started working on the issue. For example, the bug report might have remained unattended for a long period before any action was taken.

To address this limitation, we also explored the time from the bug report's first comment time to the resolution time. This approach provides a more accurate measure of the active effort and time developers dedicated to resolving the bug, which means the bug report has been noticed. 

Both metrics complement each other, which can provide an overview of the manual effort involved by developers in bug resolution.

\subsubsection*{\textbf{Results}}

Table~\ref{tab:table_Jira_results} presents the results of our analysis of 127 privacy-related bug reports. 
On average, it takes 175.31 days from the creation of a bug report to its resolution, and 140.86 days from the first comment to the resolution, which indicates that resolving bug reports typically requires a significant amount of time and considerable manual effort.
The standard deviation is much higher than the average, indicating a right-skewed distribution of resolution times, with a significant number of bugs that take much longer to resolve.  This also reflects the complexity and challenges developers encounter when handling these bugs.
The maximum resolution time is 2,309 days, indicating that, in some extreme cases, it can take several years to resolve a bug report.


\begin{table}[ht]
\small
\centering
\caption{An overview of the resolution times for bug reports.} \label{tab:table_Jira_results}    
\begin{threeparttable}
\begin{tabular}{@{}lrr@{}}
\toprule

\textbf{Metric} & \textbf{Time-1} & \textbf{Time-2}\\
\midrule
Minimum     & 0	& 0 \\
Maximum    & 2,309	& 2,309 \\
Average     &  175.31      & 140.86\\
Standard Deviation  &  344.46      & 312.54\\
\bottomrule
\end{tabular}
\end{threeparttable}
\end{table}
\vspace{-0.35em}
\noindent
\begin{minipage}{0.95\linewidth}
\footnotesize
\textbf{Notes:} \textit{Time-1} is the number of days from the creation time of the bug report to its resolution time, and \textit{Time-2} is the number of days from the first comment time to the resolution time.
\end{minipage}
\vspace{0.7em}

Based on the analysis in Table~\ref{tab:table_Jira_results}, we can conclude that, in general, developers need to invest significant effort to resolve bugs. However, the large standard deviation raises concerns about the uneven distribution of resolution times. Therefore, we need a more fine-grained study to investigate developers' manual effort in bug resolution. Specifically, we aim to explore whether the priority levels of bug reports influence the amount of manual effort required from developers.


In Apache Jira, a bug report can be assigned a priority level that reflects its importance~\footnote{\url{https://issues.apache.org/jira/ShowConstantsHelp.jspa}}. The priority level is manually determined by individuals who have a deep understanding of the bug's severity and urgency, such as the bug reporters, project managers, and other key stakeholders. Priority levels typically have a significant impact on the order in which developers address bug fixes.
As shown in Table~\ref{tab:labels}, bug reports in Apache Jira are categorized into five priority levels, ranging from \textit{Blocker} to \textit{Trivial} (from the most severe to the least). 

\begin{table}[ht]
\small
\centering
\caption{Bug report priority levels} \label{tab:labels}
\begin{threeparttable}
\begin{tabular}{@{}lp{7cm}@{}}
\toprule
\textbf{Priority} & \textbf{Description}\\
\midrule
Blocker & Issues that block development or testing work, preventing production from running. 
\\
\midrule
Major & Significant loss of functionality, often treated as the default priority (P2) and addressed based on community practices. 
\\
\midrule
Critical & Crashes, loss of data, or severe memory leaks. 
\\
\midrule
Minor & Minor loss of function or other issues with an easy workaround. 
\\
\midrule
Trivial & Cosmetic problem like misspelled words or misaligned text. \\
\bottomrule
\end{tabular}
\end{threeparttable}
\end{table}


Table~\ref{tab:table_priority} presents the distribution of priority levels and the corresponding developer time spent on bug resolution. Based on the table, we observe that trivial bugs are often resolved very quickly. Although they are assigned the lowest priority, these bugs are typically simple and easy to fix. As a result, developers may choose to address them promptly to maintain high productivity. This observation indicates that priority levels are not the only factor influencing developers' bug resolution time, and bug complexity also plays a significant role, particularly for simple bugs.
Bug reports without assigned priority levels take the longest time to be resolved, which indicates the significant impact of priority levels on resolution time.
Additionally, we observe that blocker bugs are resolved significantly quicker than other severe bugs. This suggests that when a bug is severe enough to block the entire software development process, developers prefer to invest greater effort to ensure it is resolved as quickly as possible. In such cases, the complexity of the bug becomes less of a determining factor, and developers prioritize resolving the severe bugs.


\begin{table}[ht]
\small
\centering
\caption{Overview of bug report priority levels and their resolution time.} \label{tab:table_priority}    
\begin{threeparttable}
\begin{tabular}{@{}lrrr@{}}
\toprule

\textbf{Priority} & \textbf{Count} &  \textbf{Time-1} & \textbf{Time-2}\\
\midrule
Blocker   & 7 & 76	& 112.98 \\
Major \& P2    & 86 &  169.52      & 143.87\\
Critical  & 15  & 229.07	& 148.87 \\
Minor   & 14 & 229.36	& 142.57 \\
Trivial   & 2 & 13	& 8.00 \\
None   & 3 & 160.33	& 160.33 \\

\midrule    

\textbf{Total} & 127 & 175.31 & 140.86 \\
\bottomrule
\end{tabular}
\end{threeparttable}
\end{table}






To examine the relationship between bug report priority levels and the time taken to resolve it (Time-1 and Time-2), we calculated the Spearman correlation coefficient between priority and resolution time. 
Spearman correlation coefficient is a commonly used statistical method for capturing monotonic relationships based on rank order \cite{schober2018correlation}. 
It ranges from -1 to 1, with values close to 1 indicating a strong positive monotonic relationship, values close to -1 indicating a strong negative monotonic relationship, and values near 0 suggesting no consistent trend in ranking between the variables \cite{xiao2016using}.


To calculate the Spearman correlation coefficient, we first assign a numerical value to each priority level.
Specifically, we assigned integer values ranging from -3 to 1, corresponding to priorities from lowest to highest. 
We assigned \textit{Major} a value of 0, as it is commonly treated as the baseline or default priority level (P2) in software development workflows and it is the most frequent category in the dataset (86 out of 127 cases).

Table~\ref{tab:table_spearman} presents the results of our Spearman correlation analysis.
During the analysis, we excluded three invalid bug reports that lacked priority level information.
For the remaining bug reports, we calculated the Spearman correlation coefficients between priority levels and both Time-1 and Time-2 for two different sets of bug reports.
The first set includes all 124 valid bug reports, while the second set excludes those \textit{Blocker} and \textit{Trivial} bug reports, which represent two relatively extreme cases.
The Spearman correlation coefficients indicate a weak negative monotonic relationship between bug priority and resolution time, i.e., higher priority bugs tend to be resolved more quickly.
The correlation is stronger for Time-1 than for Time-2, which indicates priority levels have a greater influence on the overall bug report resolution time, rather than on the resolution time where developers are actively focused.
After excluding \textit{Blocker} and \textit{Trivial} bugs, the absolute values of the coefficients increase (from -0.23 to -0.27 at Time-1 and from -0.12 to -0.17 at Time-2), suggesting that these types of bugs with extreme priority levels may distort the overall trend.





\begin{table}[ht]
\small
\centering
\caption{Spearman correlation coefficient between priority levels and resolution time.} \label{tab:table_spearman}    
\begin{threeparttable}
\begin{tabular}{@{}lrr@{}}
\toprule

\textbf{Data} & \textbf{Time-1} &  \textbf{Time-2}\\
\midrule
All Valid Bugs    & -0.23  & -0.12  \\
w/o Blocker \& Trivial  &  -0.27 & -0.17 \\

\bottomrule
\end{tabular}
\end{threeparttable}
\end{table}




\begin{Summary}{Summary of PQ}{}
Resolving bug reports can require significant manual effort and be very time-consuming, with an average resolution time of 175.31 days and 140.86 days from the first comment to resolution. Bugs with higher priority levels tend to be resolved more quickly.
\end{Summary}

\subsection*{\textbf{RQ1: How are the basic capabilities of LLMs in decomposing bug reports?}}

\subsubsection*{\textbf{Motivation}}


As mentioned in Section~\ref{sec:background}, bug reports are often complex and contain diverse types of information, which makes effective analysis challenging.
With the emergence of LLMs, their powerful reasoning capabilities and natural language understanding have made the application of LLMs in software development activities increasingly common.
Therefore, our objective is to explore the potential of LLMs to facilitate bug report analysis, in particular, to determine whether LLMs can be used to decompose bug reports into smaller, self-contained units.


\subsubsection*{\textbf{Approach}}






We choose the most commonly used LLMs and automate the bug report decomposition process via scripts, which consists of the following steps:

\begin{itemize}
    \item \textit{Prompt design.} Since we aim to study the basic capabilities of LLMs for bug report decomposition, we deliberately avoid using any advanced prompt engineering techniques in this RQ. Instead, we provide the LLMs with plain and straightforward prompts (zero-shot prompting) that clearly express our intention: to decompose bug reports into smaller, self-contained units. Each prompt also includes a description of the bug report to be decomposed.
This approach differs from traditional bug report analysis tasks, which typically focus on labeling or classifying bug reports~\cite{Nachai2014, zhang2015survey}.
To avoid interference between different prompt inquiries, we clear the conversation history after each one and start a fresh session.

    \item \textit{Manual Analysis.} To assess the accuracy and reliability of the bugs identified by the LLMs, we conduct a manual analysis of the decomposition results. 
The first two authors independently review all decomposed bug reports and check whether they are accurate, self-contained, and aligned with the original bug reports.
The checking decision involves two key considerations: (1) whether the number of decomposed bugs is correct, and (2) whether the description of each decomposed bug is accurate.
For each bug report, we not only evaluate whether the LLMs successfully decompose it, but also perform an in-depth analysis of the unsuccessful cases to study why the LLMs failed to produce a successful decomposition.
The reviewers then merge their review results after discussion. For disagreement results, a third person acts as a mediator
until an agreement is reached.

\end{itemize}

\subsubsection*{\textbf{Results}}

We adopt \texttt{GPT-4o} and \texttt{DeepSeek-V1} as the two LLMs for analysis, and the reasons for this choice are explained in Section~\ref{sec:setup}.
Table \ref{tab:table_rq1_LLM_T&F_detail} displays the decomposition results of each LLM. 
In general, the basic capabilities of LLMs for bug report decomposition are insufficient. After manual analysis, only 10 out of 127 bug reports were correctly decomposed by ChatGPT, and 11 by DeepSeek. 
A detailed comparison between these two LLMs can be found in Section~\ref{sec:discussion}.



Through an in-depth analysis of the incorrectly decomposed cases, we identified four major reasons:



\begin{table}[H]
\small
\centering
\caption{A summary of bug report decompositions by different LLMs.} \label{tab:table_rq1_LLM_T&F_detail}    
\begin{tabular}{@{}llrr@{}}
\toprule

\textbf{Label} & \textbf{Rationale} & \textbf{ChatGPT} & \textbf{DeepSeek}\\
\midrule
True        &  / & 10 & 11 \\
\midrule
False	    & Over-decomposition &  75 & 72 \\
False	    & Over-analysis &  31 & 36 \\
False	    & Incorr. sol. interp. &  9 & 3\\
False	    & Lacking key information  & 2  & 5\\
\midrule
\textbf{False Total} & & 117 & 116 \\

\midrule
\textbf{Total} & 	& 127 & 127 \\                
\bottomrule
\end{tabular}
\end{table}
\vspace{-0.35em}
\noindent
\begin{minipage}{0.95\linewidth}
\footnotesize
\textbf{Notes:} \textit{True} refers to cases where the LLMs accurately decompose the bug report, while \textit{False} refers to cases where the LLMs fail to decompose the bug report accurately.
\end{minipage}

\vspace{0.7em}

 \paragraph{Over-decomposition} 
 LLMs often over-decompose bug reports, resulting in the number of decomposed bugs exceeding the expected number.
LLMs may split a single bug description into multiple interwoven sub-problems, thereby making it more difficult for developers to resolve the bugs and identify their root causes. These decomposed bugs may use different wording but often refer to the same underlying issue.

For example, in the case below (cf. Table~\ref{tab:over-decomposition}), ChatGPT incorrectly decomposes the bug description into two smaller units, both of which involve the same scenario. This bug description should not have been decomposed. LLMs need a better understanding of the bug scenarios to avoid such over-decomposition.

\begin{table}[ht]
\small
\centering
\caption{An example of over-decomposition by ChatGPT.}
\label{tab:over-decomposition}    
\begin{tabular}{@{}|p{8.4cm}|@{}}
\hline
\cellcolor{orange!15} \textbf{Bug description:} \\ \hline
When RGB bitmaps are used, there can be performance improvements made to mitigate some of the impact.
\\
\hline
\cellcolor{orange!15} \textbf{Decomposition result:}\\ \hline
 Using RGB bitmaps leads to performance impacts.       \\
\hline
 Implement performance improvements to mitigate the impact when RGB bitmaps are used.\\
\hline
\end{tabular}
\end{table}




Additionally, we observed that LLMs may add extra information to the decomposed bugs, even though our prompt explicitly instructed them not to do so. This additional information often leads to an unnecessary increase in the number of decomposed bugs.
In the example below (cf. Table~\ref{tab:table_rq2_deepseek}), DeepSeek incorrectly decomposes the bug report into three separate units, while adding unnecessary descriptions that are not present in the original report.
Please note that, to save space in the paper, we do not include the full bug report descriptions. Instead, we present only the example bug descriptions and their corresponding decomposed bugs.

\begin{table}[ht]
\small
\centering
\caption{An example of over-decomposition by DeepSeek.}
\label{tab:table_rq2_deepseek}    
\begin{tabular}{@{}|p{8.4cm}|@{}}
\hline
\cellcolor{orange!15} \textbf{Bug description:} \\ \hline
As the web site use google analytics privacy policy page must be added.\\
\hline
\cellcolor{orange!15} \textbf{Decomposition result:}\\ \hline
 Add a privacy policy page to the website.\\
\hline
 Integrate Google Analytics tracking on the website.\\ 
\hline
 Ensure the privacy policy page complies with Google Analytics requirements.\\
\hline
\end{tabular}
\end{table}

\paragraph{Over-analysis} We observed that LLMs tend to analyze the content of the bug report, such as logs or code, and even attempt to offer solutions based on the analysis of bugs, which is not consistent with our goal. 


In our prompts, we explicitly stated that we were only interested in the decomposition of bug reports, and emphasized that the decomposed bugs should be kept as close to their original form as possible. However, we still find that, in some cases, the solutions produced by LLMs exist, regardless of their correctness, requiring additional manual verification and increasing the overall complexity of the bug resolution process.

In the example shown in Table~\ref{tab:table_rq2_overanalysis_gpt}, the original bug report includes logs of an exception's stack trace. When ChatGPT decomposes the bug report, it analyzes each log line individually, and creates a sub-bug description for each one. However, this bug description is specific to an I/O exception that occurred while parsing a PDF. 
This over-analysis led to an unexpected increase in the number of decomposed bugs.













\begin{table}[ht]
\small
\centering
\caption{An example of over-analysis by ChatGPT.}
\label{tab:table_rq2_overanalysis_gpt}    
\begin{tabular}{@{}|p{8.4cm}|@{}}
\hline
\cellcolor{orange!15} \textbf{Bug description:} \\ \hline
Oleg Krechowetzki reported an issue with the non sequential parser via private mail. He provided a working solution and a test pdf which can't be attached due to privacy reasons. 

The following exception occurs when parsing the pdf in question using the non sequential parser: 

\texttt{\scriptsize Caused by: java.io.IOException: Error: Expected a long type, actual='xref' }\\
\texttt{\scriptsize at org.apache.pdfbox.pdfparser.BaseParser.readLong(}\\
\texttt{\scriptsize BaseParser.java:1668)}\\ 
\texttt{\scriptsize at org.apache.pdfbox.pdfparser.BaseParser.}\\
\texttt{\scriptsize readObjectNumber(BaseParser.java:1598)}\\
\texttt{\scriptsize at org.apache.pdfbox.pdfparser.NonSequentialPDFParser.}\\
\texttt{\scriptsize parseXrefObjStream(NonSequentialPDFParser.java:458)}

\\
\hline
\cellcolor{orange!15} \textbf{Decomposition result:}\\ \hline
    Oleg Krechowetzki reported an issue with the non sequential parser via private mail. \\ \hline

 The provided test pdf cannot be attached due to privacy reasons. \\ \hline

An exception occurs when parsing the pdf in question using the non sequential parser. \\ \hline

 An exception occurs when parsing the pdf in question using the non sequential parser. The exception is a \texttt{\scriptsize java.io.IOException} with the message: ``Error: Expected a long type, actual=`xref'''. \\ \hline

  An exception occurs when parsing the pdf in question using the non sequential parser. The exception originates from the method \texttt{\scriptsize org.apache.pdfbox.pdfparser.BaseParser.readLong(BaseParser.}\\\texttt{\scriptsize java:1668)}. \\ \hline

 An exception occurs when parsing the pdf in question using the non sequential parser. The exception is also related to the method \texttt{\scriptsize org.apache.pdfbox.pdfparser.BaseParser.readObjectNumber(.}\\\texttt{\scriptsize BaseParser.java:1598)}. \\ \hline

 An exception occurs when parsing the pdf in question using the non sequential parser. The exception is triggered in the method \texttt{\scriptsize org.apache.pdfbox.pdfparser.NonSequentialPDFParser.}\\\texttt{\scriptsize parseXrefObjStream(NonSequentialPDFParser.java:458)}. \\
\hline
\end{tabular}
\end{table}










\paragraph{Incorrect Interpretation of Solutions} 
In this case, LLMs incorrectly classified the solutions suggested by reporters in bug reports. Ideally, when decomposing these bug reports, the LLM should associate each suggested solution with the corresponding bug it aims to address, rather than treating the solution as a separate bug. This misclassification not only leads to an unexpected increase in the number of decomposed bugs, but also results in a mismatch between suggested solutions and the corresponding bugs, thereby increasing the complexity of bug resolution for developers.
Some suggested solutions may include code snippets that developers use to trace the bug location. However, LLMs tend to interpret these code snippets instead of preserving the original form of the solutions. This behavior not only increases the complexity of bug decomposition but may also introduce misleading guidance for developers.

In the example below (cf. Table~\ref{tab:rq2_incor_sol}), ChatGPT and DeepSeek misclassify a suggested solution, \textit{restarting the Kylin server}, as an isolated unit, rather than linking it to the two related bugs it aims to address. This misclassification prevents the suggested solution from receiving proper attention from developers and undermines effective bug resolution.






\begin{table}[ht]
\small
\centering
\caption{An example of incorrect Interpretation of Solutions by ChatGPT and DeepSeek}
\label{tab:rq2_incor_sol}    
\begin{tabular}{@{}|p{8.4cm}|@{}}
\hline
\cellcolor{orange!15} \textbf{Bug description:} \\ \hline
There are two cases of query failure when calcite parses sql. However, both are solved by the same solution, just restarting kylin server.\\
\hline
\cellcolor{orange!15} \textbf{ChatGPT's Decomposition:}\\ \hline
 There is a query failure case when calcite parses SQL.\\ \hline
 There is a second query failure case when calcite parses SQL.\\  \hline
 Both query failures can be resolved by restarting the Kylin server.\\ 
\hline
\cellcolor{orange!15} \textbf{DeepSeek's Decomposition:}\\ \hline
Query failure occurs when Calcite parses SQL in the first case.\\ \hline
 Query failure occurs when Calcite parses SQL in the second case.\\ \hline
 Restarting the Kylin server resolves both query failure cases.\\ 
\hline
\end{tabular}
\end{table}

\paragraph{Lacking key information} 
In this case, LLMs generate an appropriate number of decomposed bugs but omit key information from the original bug reports that could aid in bug resolution. This includes information such as reproducing code, logs, and reference links provided by developers. As a result, developers still need to refer to the original bug reports to retrieve the missing information. In the following example (cf. Table~\ref{tab:rq1_lack_detail}), although ChatGPT successfully decomposed the bug description into three distinct bugs, it omitted the reference URL provided by the developer, which could have facilitated the resolution process.






\begin{table}[ht]
\small
\centering
\caption{An example of Lacking key information by ChatGPT.}
\label{tab:rq1_lack_detail}    
\begin{tabular}{@{}|p{8.4cm}|@{}}
\hline
\cellcolor{orange!15} \textbf{Bug description:} \\ \hline
As per ASF privacy policy [1], we should not host JS or CSS files on CDNs such as Cloudflare instead, we should opt to host the Fonts/JS/CSS locally on the Apache server. 

We have multiple instances [2] where we are downloading scripts from cloudflare. This task should remove this dependency and host the files locally. 

[1] https://privacy.apache.org/faq/committers.html 
[2] Instances of Fonts/JS/CSS downloaded from CDN (not exhaustive): 
\textcolor{red}{https://github.com/apache/tinkerpop/blob/master/docs/site/home/}\\\textcolor{red}{index.html\#L26 } \\
\textcolor{red}{https://github.com/apache/tinkerpop/blob/master/docs/site/home/}\\ \textcolor{red}{index.html\#L951}
\\
\hline
\cellcolor{orange!15} \textbf{Decomposition result:}\\ \hline
Remove the dependency on Cloudflare for hosting JS files and host them locally on the Apache server.\\ \hline
Remove the dependency on Cloudflare for hosting CSS files and host them locally on the Apache server.\\ \hline
Remove the dependency on Cloudflare for hosting Fonts and host them locally on the Apache server. \\
\hline
\end{tabular}
\end{table}


\begin{Summary}{Summary of RQ2}{}
Both ChatGPT and DeepSeek show insufficient basic capabilities in bug decomposition, with only 10 and 11 out of 127 bug reports correctly decomposed, respectively. We identified four major reasons for incorrect decomposition: \textit{over-decomposition}, \textit{over-analysis}, \textit{incorrect interpretation of solutions}, and \textit{lacking key information}, with \textit{over-decomposition} being the most significant reason.
\end{Summary}

\subsection*{\textbf{RQ2: How effectively do LLMs decompose bug reports using few-shot?}}

\subsubsection*{\textbf{Motivation}}


In RQ1, we explored the basic capabilities of LLMs in bug decomposition, but the results were not promising. Therefore, for this RQ, we aim to explore the potential for improving LLMs' bug decomposition abilities. The performance of existing LLMs in handling software engineering tasks is influenced by the quality of prompts~\cite{white2024chatgpt}, which has inspired us to investigate whether we can incorporate advanced prompt engineering techniques to enhance LLMs' bug decomposition capabilities. It is widely known that LLMs have strong learning abilities, which enables them to learn from successful cases and apply what they have learned to unexplored cases. The few-shot learning technique within prompt engineering aligns perfectly with this well-known concept, making the application of prompt engineering in bug decomposition practical.

\subsubsection*{\textbf{Approach}}
Similar to RQ1, the approach of this RQ consists of two steps: prompt design and manual analysis.
The major difference between the approaches for RQ1 and RQ2 is that, for RQ2, we adopt a few-shot approach when designing prompts, i.e., the prompts need to include several examples of successful bug decompositions as demonstrations. Below, we explain how we select and present these demonstrations.



Our demonstration selection was subjected to the following subtle considerations: (1) The length of bug description should be appropriate, not too long, to avoid excessive and irrelevant information that could increase the difficulty for LLMs to understand. (2) The decomposition approach should be straightforward and easy for LLMs to learn. (3) Ideally, the decomposition should cover different types, which could enable LLMs to handle a wide variety of bug reports. Eventually, we selected three bug decomposition examples as demonstrations. The decomposed bug results are based on prior LLM results and our manual enhancements.

\begin{itemize}
    \item \textit{The straightforward case:} Below is an example (cf. Table~\ref{tab:straight_demo}) of a straightforward decomposition. The decomposition does not require a deep understanding of the bug report, as it can be based on the structure of the bug report. After the sentence ``The Kafka website has a few issues, including:'', each paragraph can be treated as a separate bug, except for the last one. This straightforward decomposition provides LLMs with clear guidance on how to decompose bug reports.

\begin{table}[ht]
\small
\centering
\caption{An example of straightforward demonstration.}
\label{tab:straight_demo}    
\begin{tabular}{@{}|p{8.4cm}|@{}}
\hline
\cellcolor{orange!15} \textbf{Bug description:} \\ \hline
The ASF has updated its privacy policy and all websites must be compliant.\\
The full guidelines can be found in https://privacy.apache.org/faq/\\committers.html \\
The Kafka website has a few issues, including:\\
It's missing a link to the privacy policy: https://privacy.apache.org/\\policies/privacy-policy-public.html\\
It's using Google Analytics\\
It's using Google Fonts\\
It's using scripts hosted on Cloudflare CDN\\
Embedded videos don't have an image placeholder\\
As per the email sent to the PMC, all updates have to be done by July 22. \\
\hline
\cellcolor{orange!15} \textbf{Decomposition result:}\\ \hline
The Kafka website missing a link to the privacy policy: https://privacy.apache.org/policies/privacy-policy-public.html.\\ \hline
The Kafka website is using Google Analytics.\\ \hline
The Kafka website is using Google Fonts.\\ \hline
The Kafka website is using scripts hosted on Cloudflare CDN.\\ \hline
The Kafka website has embedded videos that don't have an image placeholder.\\
\hline
\end{tabular}
\end{table}

\item \textit{The case where code/logs require special attention to avoid over-analysis.} In RQ1, we observed that when a bug report includes code or logs, LLMs tend to over-analyze them, which leads to unnecessarily complex decomposed bugs and an unexpectedly high number of them. To avoid this, we provided a demonstration that includes code snippets (see Table~\ref{tab:code_handling}). These code snippets should serve as supplementary information for the decomposed bug descriptions and should be preserved in their original form, rather than being summarized or analyzed.

\begin{table}[ht]
\small
\centering
\caption{An example of special handling for code in a bug report.}
\label{tab:code_handling}    
\begin{tabular}{@{}|p{8.4cm}|@{}}
\hline
\cellcolor{orange!15} \textbf{Bug description:} \\ \hline
When the generated stub processes response, it expects the nodes come in a specific order. If the order is different, exception is thrown.
The code is in parse() methods of different result processing factory.
For example, the code:\\
\textcolor{blue}{
\texttt{ \scriptsize // process the "first" element "IsPrivacyChanged" } }\\
\textcolor{blue}{ \texttt{ \scriptsize if (reader.isStartElement() ... } }\\
\textcolor{blue}{ \texttt{ \scriptsize //now it processes 2nd element. }} \\
\textcolor{blue}{ \texttt{ \scriptsize if (reader.isStartElement() ... }}

If ``Name'' came in before ``IsPrivacyChanged'', exception is thrown.

A loop should be used, and any elements can appear in any order. \\
\hline
\cellcolor{orange!15} \textbf{Decomposition result:}\\ \hline

When the generated stub processes a response, it currently expects nodes to come in a specific order, and throws an exception if the order differs. If "Name" came in before "IsPrivacyChanged", exception is thrown.
In the parse() methods of different result processing factories. A loop should be used, and any elements can appear in any order.The following is the code provided by descriptions:\\
\textcolor{blue}{
\texttt{ \scriptsize // process the "first" element "IsPrivacyChanged" }}\\
\textcolor{blue}{ \texttt{ \scriptsize if (reader.isStartElement() ... }}\\
\textcolor{blue}{ \texttt{ \scriptsize  //now it processes 2nd element.}} \\
\textcolor{blue}{ \texttt{ \scriptsize if (reader.isStartElement() ... }}
\\
If ``Name'' came in before ``IsPrivacyChanged'', exception is thrown.
A loop should be used, and any elements can appear in any order. \\
\hline
\end{tabular}
\end{table}

\item \textit{ The case where the suggested solution in a bug report requires special attention: } In RQ1, we found that the suggested solution in a bug report may lead LLMs to perform extra analysis, even though we instructed them not to do so. 
To avoid this, we provided a decomposition example as a demonstration in this RQ. In the example below (cf. Table~\ref{tab:sol_handling}), the bug report is decomposed into two sub-bugs: one related to the missing user context in the view query and the other concerning CouchDB's security. However, both decomposed bugs should have the same solution, which should be duplicated and appended at the end of each decomposed bug, without any additional analysis.

\begin{table}[ht]
\small
\centering
\caption{An example of special handling for the suggested solution in a bug report.}
\label{tab:sol_handling}    
\begin{tabular}{@{}|p{8.4cm}|@{}}
\hline
\cellcolor{orange!15} \textbf{Bug description:} \\ \hline
We omitted to pass user\_ctx down in fabric\_view\_all\_docs. Since auth has happened beforehand this hasn't been an obvious issue, but it matters for the \_users db as that reacts differently based on the user. couchdb intentionally hides design documents in that database from non-admins and intentionally hides the user docs of other users.
\textcolor{blue}{Passing the user ctx down could fix it.} \\
\hline
\cellcolor{orange!15} \textbf{Decomposition result:}\\ \hline
We omitted to pass user\_ctx down in fabric\_view\_all\_docs. Since auth has happened beforehand this hasn't been an obvious issue, but it matters for the \_users db as that reacts differently based on the user. \textcolor{blue}{Passing the user ctx down could fix it.}\\ \hline
Couchdb intentionally hides design documents in that database from non-admins and intentionally hides the user docs of other users. \textcolor{blue}{Passing the user ctx down could fix it.} \\
\hline
\end{tabular}
\end{table}

\end{itemize}

\subsubsection*{\textbf{Results}}

Table~\ref{tab:rq2_result} presents the bug decomposition results for ChatGPT and DeepSeek using the few-shot approach.
Unlike RQ1, the results show that LLMs exhibit improved capabilities in bug decomposition after prompt tuning, with the number of correctly decomposed cases increasing from 10 to 24 for ChatGPT, and from 11 to 29 for DeepSeek. Although the overall number of correct cases remains relatively low, this improvement offers promising insight that prompt engineering can be leveraged to enhance the bug decomposition capability of LLMs.


\begin{table}[ht]
\small
\centering
\caption{A summary of decompositions of bug reports by different LLM after prompt promotion.} \label{tab:rq2_result}    
\begin{tabular}{@{}llrr@{}}
\toprule

\textbf{Label} & \textbf{Rationale} & \textbf{ChatGPT} & \textbf{DeepSeek}\\
\midrule
True        & /  & 24 & 29\\
\midrule
False	    & Over-decomposition &  56 & 52 \\
False	    & Over-analysis &  28 & 25\\
False	    & Incorr. sol. interp. &  18 & 12\\
False	    & Lacking key information & 1  & 4\\
\midrule
\textbf{FP Total} & & 101 & 98\\
\midrule
\textbf{Total} & 	& 127 & 127 \\                
\bottomrule
\end{tabular}
\end{table}


Among the 101 false cases generated by ChatGPT, 56 were due to over-decomposition, making it the most significant cause of incorrect decomposition. Similarly, DeepSeek produced 52 over-decomposition cases out of 98 false cases. These results suggest that LLMs often generate more units than necessary during the decomposition process. In such cases, the LLMs show limitations in semantic understanding and analysis of the bug reports, as they tend to either repeat the same issue in different forms or unnecessarily split a single bug into multiple smaller units.


Additionally, the false cases related to \textit{Over-analysis} and \textit{Incorrect Interpretation of Solutions} indicate that LLMs struggle to effectively process technical content in bug reports, such as code, logs, and suggested solutions. Although we provided examples to help guide LLMs in handling such information, the improvement remained limited. We explicitly stated in the prompt not to analyze the code and logs, yet LLMs often ignored these instructions. In many cases, the LLMs either omitted this information or misinterpreted it as a separate unit. These limitations suggest that, for developers to achieve better bug decomposition, it is essential to improve the adherence of LLMs to instructions and prevent them from being distracted by demonstrations or verbose bug descriptions. Similarly to the over-decomposition issue, enhancing the LLMs' understanding of bug reports is also important.

\begin{Summary}{Summary of RQ2}{}
The LLMs' bug decomposition capabilities are improved by using few-shot prompting, but there is still considerable room for improvement, with over-decomposition being the primary cause of errors. Better results could be achieved by enhancing the LLMs' understanding of bug descriptions and improving their adherence to prompt instructions.
\end{Summary}

\section{Discussion}\label{sec:discussion}

In this section, we discuss the results from three perspectives.



\noindent \textbf{\textit{D1: What defines a high-quality bug report?}}


Based on our analysis of bug reports, we provide the following three useful suggestions for bug reporters.


\begin{itemize}
    \item \textit{Atomic bug reports.} Our study shows that the diversity and complexity of bug report content pose significant challenges to LLMs’ bug decomposition capabilities. Therefore, we recommend that bug reporters simplify their submissions as much as possible. When a report includes multiple isolated issues, it is advisable to submit them as separate bug reports rather than merging them into one.
    An example below (cf. Table~\ref{tab:dis_complex_bug}) includes two separate bugs: an incorrect link and 108 unresolved errors. Since these issues are unrelated, combining them in a single report increases the complexity of bug resolution for developers and reduces the possibility of timely fixes. A better practice is to submit them as separate bug reports.

\begin{table}[ht]
\small
\caption{An example of a complex bug report.}
\centering
\label{tab:dis_complex_bug}    
\begin{tabular}{@{}|p{8.4cm}|@{}}
\hline
\cellcolor{orange!15} \textbf{Bug description:} \\ \hline
The current webpage at https://creadur.apache.org/rat/ links to mail-lists.html, which should be https://creadur.apache.org/rat/\\mailing-lists.html \\
Check for other dead links and fix them appropriately.\\ 
Further analysis via \{\{\$ linkchecker https://creadur.apache.org/\\rat/ -F text/UTF-8/rat-linkchecker.txt  \}\} showed 108 errors on the current RAT webpage.
\\
\hline
\end{tabular}
\end{table}

    \item \textit{Well-structured bug reports. }
    Well-structured bug reports can significantly enhance bug resolution efficiency, while poorly structured bug reports can lead to misunderstandings and longer resolution times.
LLMs heavily rely on textual structure and formatting cues, such as paragraph breaks, to identify and decompose bugs. When multiple issues are embedded in a single paragraph, LLMs often struggle to determine clear boundaries between them. 
    In the following example (cf. Table~\ref{tab:dis_poorly}), the bug reporter uses ``The first issue'' to indicate the first issue, but fails to mark where the next one begins. In addition, ``Also'' is not a clear indicator for separating bugs. We suggest that bug reports should be well-structured, and when multiple issues are included in a bug report, they should be placed in separate paragraphs to help LLMs and developers better identify and resolve each issue.

\begin{table}[ht]
\small
\centering
\caption{An example of a poorly structured bug report.}
\label{tab:dis_poorly}    
\begin{tabular}{@{}|p{8.4cm}|@{}}
\hline
\cellcolor{orange!15} \textbf{Bug description:} \\ \hline
Notifications of new comments are sent with the commenter's email address as the FROM address.

Having the poster's address as the FROM address seems at first glance to be a ``good thing'', however it means that any returns such as mailbox full, out of office messages etc. go back to that "untrusted" address. \textbf{\textcolor{red}{Also}}, if the address is a known email spammer, the comment will be posted but the weblog owner may not get the notification. \textbf{\textcolor{red}{The first issue}} is a bit of a privacy issue, since a site may decide not to publish contributors' email addresses, only to have the sent out through this route to unknow 3rd parties.
\\
\hline
\end{tabular}
\end{table}

    \item \textit{Concise and effective bug descriptions.} In the previous RQs, we observed that when bug reports contain various types of information, such as code or logs, they can pose challenges for LLMs in bug decomposition. While such information can be useful in practice, overly verbose bug reports may hinder the bug resolution process. Therefore, it is crucial to maintain a balance by keeping bug reports concise while ensuring they include all necessary information for effective bug resolution.
\end{itemize}

\noindent \textbf{\textit{D2: Which LLM performs more effectively?}}


Table \ref{tab:table_dis_compare} presents a comparison of bug decomposition results for ChatGPT and DeepSeek under two different prompting strategies.  The results indicate no major performance differences between the two LLMs. ChatGPT slightly outperforms DeepSeek in RQ1, whereas DeepSeek slightly outperforms ChatGPT in RQ2.

\begin{table}[ht]
\small
\centering
\caption{A summary of different LLMs' performance.} \label{tab:table_dis_compare}    
\begin{tabular}{@{}lrr@{}}
\toprule
\textbf{LLM} & \textbf{True} & \textbf{False}\\
\midrule
ChatGPT (RQ1)        & 10  & 117 \\
ChatGPT (RQ2)        & 24 & 103 \\
Promotion (\%) & 240\%    & 88.03\% \\
\midrule
DeepSeek (RQ1)	    &11  &116\\
DeepSeek (RQ2)	    &29   &98 \\
Promotion (\%) & 263.64\%	& 84.48\% \\                
\bottomrule
\end{tabular}
\end{table}

\noindent \textbf{\textit{D3: Do bug resolution time and priority levels potentially impact the LLMs' capabilities to decompose bugs?}}

\begin{itemize}
    \item \textit{Relationship between bug resolution time and the LLMs' capabilities:} According to Table~\ref{tab:time_label}, bug reports that were correctly decomposed by the LLMs exhibited shorter resolution times (statistically significant) compared to those that were incorrectly decomposed. This difference is more pronounced in the case of DeepSeek than in ChatGPT, suggesting that LLMs are more likely to successfully decompose bugs that are relatively easier and quicker to resolve.

\begin{table}[ht]
\small
\centering
\caption{Average time-1 and time-2 comparison across different LLMs and final labels.} \label{tab:time_label}    
\begin{tabular}{@{}lllrr@{}}
\toprule

\textbf{Label} & \textbf{RQ}  & \textbf{LLM}  & \textbf{Avg Time-1} & \textbf{Avg Time-2}\\
\midrule
\multirow{4}{*}{True}&   \multirow{2}{*}{RQ1} & ChatGPT     & 108.90 & 137.50 \\
& & DeepSeek       & 16.27 & 33.55 \\
\cmidrule(l{1pt}r{1pt}){2-5} 
& \multirow{2}{*}{RQ2} & ChatGPT         & 166.79 & 122.99 \\
 &   & DeepSeek       & 155.03 & 110.99 \\ 
\midrule
\multirow{4}{*}{False} & \multirow{2}{*}{RQ1} & ChatGPT        & 180.99 & 141.15 \\
&   & DeepSeek     & 190.40 & 151.04 \\
\cmidrule(l{1pt}r{1pt}){2-5} 
&  \multirow{2}{*}{RQ2}   & ChatGPT       & 177.30 & 145.03 \\
&   & DeepSeek       & 181.32 & 149.70 \\            
\bottomrule
\end{tabular}
\end{table}

    \item \textit{Relationship between priority levels and the LLMs' capabilities: } According to Table~\ref{tab:priority_dis}, bug reports that were correctly decomposed by LLMs have various priority levels, with Major \& P2 being the most common. However, we cannot conclude that Major \& P2 bugs are easier to decompose, as they already account for the largest proportion of reports in the entire dataset. Therefore, there is no obvious correlation between priority levels and the LLMs’ bug decomposition capabilities.

\begin{table}[ht]
\small
\centering
\caption{Priority level distribution across LLMs, different prompting strategies, and labels.} \label{tab:priority_dis}    
\begin{tabular}{@{}llrrrr@{}}
\toprule
\multirow{2.7}{*}{\textbf{Label}} & \multirow{2.7}{*}{\textbf{Priority}} & \multicolumn{2}{c}{\textbf{RQ1}} & \multicolumn{2}{c}{\textbf{RQ2}}\\
\cmidrule(l{1pt}r{1pt}){3-4} 
\cmidrule(l{1pt}r{1pt}){5-6}
 &  & \textbf{GPT} & \textbf{DeepSeek} & \textbf{GPT}& \textbf{DeepSeek}\\
\midrule
\multirow{6}{*}{True} & Blocker  & 1 & 0 & 2 & 2 \\
 & Critical  & 0 & 2 & 3 & 4\\
& Major \& P2  & 6 & 6 & 14 & 16 \\
& Minor &  2 & 2 & 3 & 5 \\
& Trivial &   1 & 0 & 1 & 1 \\ 
&  None   &   0 & 1 & 1 & 1 \\  
\midrule
\multirow{6}{*}{False} & Blocker & 6 &  7 & 5 & 5\\
& Critical  & 15 & 13  & 12 & 11 \\
& Major \& P2   & 80 & 80 & 72 & 70 \\
& Minor & 12 & 12 & 11 & 9 \\
& Trivial & 1 & 2 & 1 & 1\\
& None  & 3 & 2 &2 & 2 \\
\bottomrule
\end{tabular}
\end{table}
    
\end{itemize}

\section{Threats to Validity} \label{sec:threat}


\noindent \textbf{Internal Validity.}
The internal validity may be subject to the following threats:
(1)
\textit{Selection of bug reports:} Our dataset is limited to 127 privacy-related bug reports from Apache Jira. This restricted scope may not fully represent the diversity of bug reports across different domains or projects, potentially limiting the generalizability of our findings, however, this focused dataset enables us to conduct a more controlled and in-depth analysis and uncover novel insight, which could serve as a foundation and inspiration for future study.
(2) \textit{Manual analysis's subjectivity:} Manual analysis may introduce subjectivity. To address this, two reviewers independently analyzed the LLMs' results, and any disagreements were resolved through careful discussion. 
(3) \textit{LLM performance variability:} The performance of LLMs can vary based on prompt phrasing and model updates. Each prompt in this study was carefully designed after thoughtful consideration, and we used the most recent LLM versions as of the time of writing, which reflects the current capabilities of state-of-the-art LLMs.

\noindent \textbf{External Validity. }
The external validity may be subject to the following threats:
(1) \textit{LLM limitations:}
The study utilized specific LLMs, such as ChatGPT and DeepSeek, for decomposing bug reports. The results may not generalize to other LLMs or AI models with different architectures, training data, or capabilities. However, these two LLMs are currently among the most popular and representative models on the market, which makes studying their performance both practical and meaningful.
(2)
\textit{Dataset specificity:}
The bug reports analyzed were sourced from the Apache Jira repository, which may not be representative of all software projects. 
Projects with different development practices, team structures, or bug tracking systems may have bug reports with diverse characteristics, which could lead to varying results in automated bug decomposition analysis.
However, Apache Jira is currently one of the most popular bug report repositories, and the projects it covers are mostly well-established. Therefore, studying bug reports from this repository can provide useful and practical insights that are relevant to many real-world software development scenarios.


\noindent \textbf{Construct Validity.}
\textit{LLM's understanding of context.}
An LLM's ability to decompose a bug report depends on its understanding of the context and technical details within the bug report. LLMs may struggle with vague or incomplete bug descriptions, resulting in inaccurate bug decomposition. Additionally, LLMs often fail to accurately understand the meaning of complex prompts or those that contain complex examples, which can result in unintended interpretations. To mitigate this, our study focuses on two of the most state-of-the-art LLMs currently available, ensuring that the findings reflect the capabilities of the most advanced LLMs in use today.

\section{conclusion} \label{sec:conclusion}

Bug reports are a critical resource in SDLC and play an important role in many key software engineering activities. However, their complexity makes automated bug report analysis challenging.
In this study, we explored the capabilities of LLMs in bug decomposition.
We conducted an empirical study on two LLMs, ChatGPT and DeepSeek, using a dataset of 127 resolved bug reports collected from the Apache Jira repository.
We found that 
prompt engineering can enhance the capabilities of LLMs in bug decomposition, but their overall capabilities still require further improvement.




In this study, we also identified four common causes of incorrect bug decompositions, including \textit{over-decomposition}, \textit{over-analysis}, \textit{incorrect interpretation of solutions}, and \textit{lacking key information}. While LLMs have strong potential in software engineering activities, they are not yet reliable enough to fully automate bug decomposition. Meanwhile, bug reporters also have a responsibility to improve the quality of their reports to facilitate automated bug analysis in the future.

In future work, we plan to expand our dataset to include more bug reports and explore additional applications of prompt engineering in automated bug report analysis.


\bibliographystyle{plainnat}
\bibliography{reference}

\end{document}